\documentclass[11pt]{article}
\usepackage{latexsym,epic,eepic}
\begin{document}
\newcommand{\2}{\vspace{0.2 cm}}
\newcommand{\dist}{{\rm dist}}
\newcommand{\diam}{{\rm diam}}
\newcommand{\rad}{{\rm rad}}
\newcommand{\dom}{\mbox{$\rightarrow$}}
\newcommand{\ldom}{\mbox{$\leftarrow$}}
\newcommand{\edom}{\mbox{$\leftrightarrow$}}
\newcommand{\qed}{\hfill$\diamond$}
\newcommand{\pf}{{\bf Proof: }}
\newtheorem{theorem}{Theorem}[section]
\newcommand{\ra}{\rangle}
\newcommand{\la}{\langle}
\newtheorem{lemma}[theorem]{Lemma}
\newtheorem{claim}[theorem]{Claim}
\newtheorem{definition}[theorem]{Definition}
\newtheorem{corollary}[theorem]{Corollary}
\newtheorem{proposition}[theorem]{Proposition}
\newtheorem{conjecture}[theorem]{Conjecture}
\newtheorem{problem}[theorem]{Problem}
\newtheorem{question}[theorem]{Question}
\newtheorem{remark}[theorem]{Remark}
\newcommand{\beq}{\begin{equation}}
\newcommand{\eeq}{\end{equation}}
\newcommand{\argmax}{{\rm argmax}}
\newcommand{\MiP}{MinHOM($H$) }
\newcommand{\MaP}{MaxHOM($H$) }
\newcommand{\vecc}[1]{\stackrel{\leftrightarrow}{#1}}

\title{Minimum Cost Homomorphisms to Semicomplete Bipartite Digraphs}

\date{}

\author{Gregory Gutin\thanks{Corresponding author. Department of Computer Science,
Royal Holloway University of London, Egham, Surrey TW20 OEX, UK,
gutin@cs.rhul.ac.uk}  \and  Arash Rafiey\thanks{School of Computing
Science, Simon Fraser University, Burnaby, B.C., Canada, V5A 1S6,
arafieyh@cs.sfu.ca} \and Anders Yeo\thanks{Department of Computer
Science, Royal Holloway University of London, Egham, Surrey TW20
OEX, UK, anders@cs.rhul.ac.uk}}

\maketitle

\begin{abstract}
For digraphs $D$ and $H$, a mapping $f:\ V(D)\dom V(H)$ is a
homomorphism of $D$ to $H$ if $uv\in A(D)$ implies $f(u)f(v)\in
A(H).$ If, moreover, each vertex $u \in V(D)$ is associated with
costs $c_i(u), i \in V(H)$, then the cost of the homomorphism $f$ is
$\sum_{u\in V(D)}c_{f(u)}(u)$. For each fixed digraph $H$, we have
the {\em minimum cost homomorphism problem for} $H$. The problem is
to decide, for an input graph $D$ with costs $c_i(u),$ $u \in V(D),
i\in V(H)$, whether there exists a homomorphism of $D$ to $H$ and,
if one exists, to find one of minimum cost. Minimum cost
homomorphism problems encompass (or are related to) many well
studied optimization problems. We describe a dichotomy of the
minimum cost homomorphism problem for semicomplete multipartite
digraphs $H$. This solves an open problem from an earlier paper. To
obtain the dichotomy of this paper, we introduce and study a new
notion, a $k$-Min-Max ordering of digraphs.
\end{abstract}

\section{Introduction}\label{introsec}

\noindent{\bf Motivation.} We consider only directed (undirected)
graphs that have neither loops nor multiple arcs (edges). In this
paper we solve a problem raised in \cite{gutinDO} to find a
dichotomy for the computational complexity of minimum cost
homomorphism problem (MCH) for semicomplete bipartite digraphs (we
define this problem below). In fact, our result leads to a complete
dichotomy for the computational complexity of MCH for semicomplete
$k$-partite digraphs ($k\ge 2$) as a (much simpler) dichotomy for
the case $k\ge 3$ was obtained in \cite{gutinDO} (see also Section
\ref{gdsec}). Our result uses and significantly extends a dichotomy
for the computational complexity of MCH for bipartite undirected
graphs obtained in \cite{gutinEJC}.

In our previous papers we used properties of an important notion of
Min-Max ordering of digraphs. To obtain the dichotomy of this paper,
we introduce and study a new notion, a $k$-Min-Max ordering of
digraphs. We believe that properties of this notion and, in
particular, Theorem \ref{mm2} can be used to obtain further results
on MCH and its special cases (see below).

The minimum cost homomorphism problem was introduced in
\cite{gutinDAMlora}, where it was motivated by a real-world problem
in defence logistics. We believe it offers a practical and natural
model for optimization of weighted homomorphisms. MCH's special
cases include the well-known list homomorphism problem
\cite{hell2003,hell2004} and the general optimum cost chromatic
partition problem, which has been intensively studied
\cite{halld2001,jansenJA34,jiangGT32}, and has a number of
applications, \cite{kroon1997,supowitCAD6}.

\noindent{\bf Minimum cost homomorphisms.} For directed or
undirected graphs $G$ and $H$, a mapping $f:\ V(G)\dom V(H)$ is a
{\em homomorphism of $G$ to $H$} if $uv\in E(G)$ implies
$f(u)f(v)\in E(H).$ Recent treatments of homomorphisms in directed
and undirected graphs can be found in \cite{hell2003,hell2004}. Let
$H$ be a fixed directed or undirected graph. The {\em homomorphism
problem} for $H$ asks whether a directed or undirected input graph
$G$ admits a homomorphism to $H.$ The {\em list homomorphism
problem} for $H$ asks whether a directed or undirected input graph
$G$ with lists (sets) $L_u \subseteq V(H), u \in V(G)$ admits a
homomorphism $f$ to $H$ in which $f(u) \in L_u$ for each $u \in
V(G)$.

Suppose $G$ and $H$ are directed (or undirected) graphs, and
$c_i(u)$, $u\in V(G)$, $i\in V(H)$ are nonnegative {\em costs}. The
{\em cost of a homomorphism} $f$ of $G$ to $H$ is $\sum_{u\in
V(G)}c_{f(u)}(u)$. If $H$ is fixed, the {\em minimum cost
homomorphism problem}, MinHOM($H$), for $H$ is the following
optimization problem. Given an input graph $G$, together with costs
$c_i(u)$, $u\in V(G)$, $i\in V(H)$,  we wish to find a minimum cost
homomorphism of $G$ to $H$, or state that none exists.

A bipartite digraph is {\em semicomplete} if there is at least one
arc between every two vertices belonging to different partite sets.
In this paper, we study the {\em minimum cost homomorphism problem
for semicomplete bipartite digraphs}, i.e., MinHOM($H$) when $H$ is
a {\em semicomplete bipartite digraph}. Observe that MCH for
semicomplete bipartite digraphs extends MCH for bipartite undirected
graphs. Indeed, let $B$ be a semicomplete bipartite digraph with
partite sets $U,V$ and arc set $A(B)=A_1\cup A_2$, where $A_1=\{uv:\
u\in U,\ v\in V\}$ and $A_2\subseteq \{vu:\ v\in V,\ u\in U\}$. Let
$B'$ be a bipartite graph with partite sets $U,V$ and edge set
$E(B')=\{uv:\ vu\in A_2\}.$ Notice that MinHOMP($B$) is equivalent
to MinHOMP($B'$).

\noindent{\bf Min-Max ordering.} Let $H$ be a digraph. We say that
an ordering $v_1,v_2, \ldots, v_p$ of $V(H)$ is a {\em Min-Max
ordering of $H$} if $v_iv_r, v_jv_s \in A(H)$ implies
$v_{\min\{i,j\}}v_{\min\{s,r\}} \in A(H)$ and
$v_{\max\{i,j\}}v_{\max\{s,r\}} \in A(H)$. One can easily see that
$v_1,v_2, \ldots, v_p$ of $V(H)$ is a Min-Max ordering of $H$ if
 $i < j$, $s < r$ and $v_iv_r, v_jv_s \in A(H)$, then $v_iv_s \in
A(H)$ and $v_jv_r \in A(H)$. We can define a Min-Max ordering for a
bipartite undirected graph $G$ with partite sets $V$ and $U$ as
follows: We orient all edges from $V$ to $U$ and apply the above
definition for digraphs. Importance of Min-Max ordering for
MinHOM($H$) is indicated in the following two theorems.

\begin{theorem}\label{mm1}\cite{gutinDAM}
Let a digraph $H$ have a Min-Max ordering. Then MinHOM$(H)$ is
polynomial-time solvable.
\end{theorem}

A bipartite graph $H$ with vertices $x_1,x_2,x_3,x_4,y_1,y_2,y_3$
is called
\begin{description}
\item{\em a bipartite claw } if its edge set
$E(H)=\{x_4y_1,y_1x_1,x_4y_2,y_2x_2, x_4y_3, y_3x_3\};$

\item{\em a bipartite net } if its edge set
$E(H)=\{x_1y_1,y_1x_3,y_1x_4,x_3y_2,x_4y_2,y_2x_2,y_3x_4\};$

\item{\em a bipartite tent } if its edge set
$E(H)=\{x_1y_1,y_1x_3,y_1x_4,x_3y_2,x_4y_2,y_2x_2,y_3x_4\}.$
\end{description}
See Figure \ref{III_graphs}.

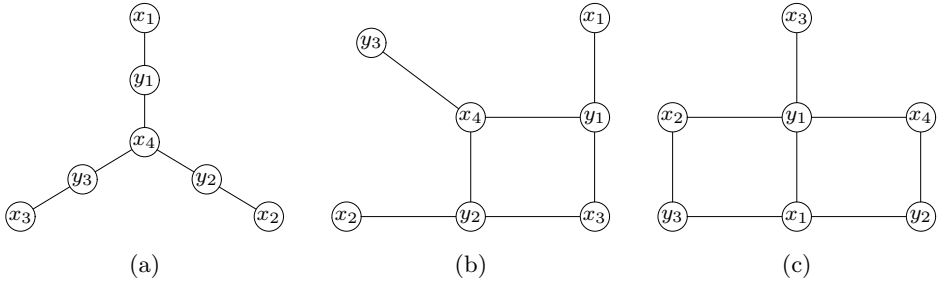
\begin{figure}
\unitlength 0.330mm \linethickness{0.4pt} \noindent
\begin{picture}(   120.00,   100.00)
\put(    60.00,    90.00){\circle{12.0}} \put(  60.000,
90.000){\makebox(0,0){{\scriptsize $x_1$}}} \put(   110.00,
10.00){\circle{12.0}} \put( 110.000,
10.000){\makebox(0,0){{\scriptsize $x_2$}}} \put(    10.00,
10.00){\circle{12.0}} \put(  10.000,
10.000){\makebox(0,0){{\scriptsize $x_3$}}} \put(    60.00,
40.00){\circle{12.0}} \put(  60.000,
40.000){\makebox(0,0){{\scriptsize $x_4$}}} \put(    60.00,
65.00){\circle{12.0}} \put(  60.000,
65.000){\makebox(0,0){{\scriptsize $y_1$}}} \put(    85.00,
25.00){\circle{12.0}} \put(  85.000,
25.000){\makebox(0,0){{\scriptsize $y_2$}}} \put(    35.00,
25.00){\circle{12.0}} \put(  35.000,
25.000){\makebox(0,0){{\scriptsize $y_3$}}} \drawline(  60.000,
84.000)(  60.000,  71.000) \drawline(  60.000,  59.000)(  60.000,
46.000) \drawline( 104.855,  13.087)(  90.145,  21.913) \drawline(
79.855,  28.087)(  65.145,  36.913) \drawline(  15.145,  13.087)(
29.855,  21.913) \drawline(  40.145,  28.087)(  54.855,  36.913)
\put(60,-10){\makebox(0,0){{\footnotesize (a)}}}
\end{picture} \hspace{-0.01cm}
\linethickness{0.4pt}
\begin{picture}(   120.00,   100.00)
\put(   110.00,    90.00){\circle{12.0}} \put( 110.000,
90.000){\makebox(0,0){{\scriptsize $x_1$}}} \put(    10.00,
10.00){\circle{12.0}} \put(  10.000,
10.000){\makebox(0,0){{\scriptsize $x_2$}}} \put(   110.00,
10.00){\circle{12.0}} \put( 110.000,
10.000){\makebox(0,0){{\scriptsize $x_3$}}} \put(    60.00,
50.00){\circle{12.0}} \put(  60.000,
50.000){\makebox(0,0){{\scriptsize $x_4$}}} \put(   110.00,
50.00){\circle{12.0}} \put( 110.000,
50.000){\makebox(0,0){{\scriptsize $y_1$}}} \put(    60.00,
10.00){\circle{12.0}} \put(  60.000,
10.000){\makebox(0,0){{\scriptsize $y_2$}}} \put(    20.00,
80.00){\circle{12.0}} \put(  20.000,
80.000){\makebox(0,0){{\scriptsize $y_3$}}} \drawline( 110.000,
84.000)( 110.000,  56.000) \drawline( 110.000,  44.000)( 110.000,
16.000) \drawline( 104.000,  10.000)(  66.000,  10.000) \drawline(
60.000,  16.000)(  60.000,  44.000) \drawline(  66.000,  50.000)(
104.000,  50.000) \drawline(  16.000,  10.000)(  54.000,  10.000)
\drawline(  24.800,  76.400)(  55.200,  53.600)
\put(60,-10){\makebox(0,0){{\footnotesize (b)}}}
\end{picture}  \hspace{-0.01cm}
\linethickness{0.4pt}
\begin{picture}(   120.00,   100.00)
\put(    60.00,    10.00){\circle{12.0}} \put(  60.000,
10.000){\makebox(0,0){{\scriptsize $x_1$}}} \put(    10.00,
50.00){\circle{12.0}} \put(  10.000,
50.000){\makebox(0,0){{\scriptsize $x_2$}}} \put(    60.00,
90.00){\circle{12.0}} \put(  60.000,
90.000){\makebox(0,0){{\scriptsize $x_3$}}} \put(   110.00,
50.00){\circle{12.0}} \put( 110.000,
50.000){\makebox(0,0){{\scriptsize $x_4$}}} \put(    60.00,
50.00){\circle{12.0}} \put(  60.000,
50.000){\makebox(0,0){{\scriptsize $y_1$}}} \put(   110.00,
10.00){\circle{12.0}} \put( 110.000,
10.000){\makebox(0,0){{\scriptsize $y_2$}}} \put(    10.00,
10.00){\circle{12.0}} \put(  10.000,
10.000){\makebox(0,0){{\scriptsize $y_3$}}} \drawline(  60.000,
16.000)(  60.000,  44.000) \drawline(  60.000,  56.000)(  60.000,
84.000) \drawline(  16.000,  50.000)(  54.000,  50.000) \drawline(
10.000,  44.000)(  10.000,  16.000) \drawline( 104.000,  50.000)(
66.000,  50.000) \drawline( 110.000,  44.000)( 110.000,  16.000)
\drawline(  54.000,  10.000)(  16.000,  10.000) \drawline( 66.000,
10.000)( 104.000,  10.000)
\put(60,-10){\makebox(0,0){{\footnotesize (c)}}}
\end{picture}

\mbox{ }

\caption{A bipartite claw  (a), a bipartite net (b) and a
bipartite tent (c).}
\end{figure}  \label{III_graphs}

\begin{theorem}\label{unddich}\cite{gutinEJC}
Let $H$ be an undirected bipartite graph. If $H$ contains a cycle
$C_{2k}$, $k\ge 3$ or a bipartite claw or a bipartite net or a
bipartite tent as an induced subgraph, then MinHOM($H$) is
NP-hard.

Assume that P$\neq$NP. Then the following three assertions are
equivalent:

(i) $H$ has a Min-Max ordering;

(ii) MinHOM($H$) is polynomial time solvable;

(iii) $H$ does not contain a cycle $C_{2k}$, $k\ge 3$, a bipartite
claw, a bipartite net, or a bipartite tent as an induced subgraph.

\end{theorem}

\noindent{\bf Additional terminology and notation.} For a graph $H$,
$V(H)$ and $E(H)$ denote its vertex and edge sets, respectively. For
a digraph $H$, $V(H)$ and $A(H)$ denote its vertex and arc sets,
respectively. For a pair $X,Y$ of vertex sets of a digraph $H$,
$(X,Y)_H$ denotes the set of all arcs of the form $xy$, where $x\in
X,y\in Y.$ We omit the subscript when it is clear from the context.
Also, $X\times Y=\{xy:\ x\in X,y\in Y\}.$ For a set $X\subseteq
V(H)$, let $N^+(X)=\{y:\ \exists x\in X \mbox{ with } xy\in A(H)\}$
and $N^-(X)=\{y:\ \exists x\in X \mbox{ with } yx\in A(H)\}.$

If $xy$ is an arc of a digraph $H$, we will say that $x$ {\em
dominates} $y$, $y$ {\em is dominated by} $x$, $y$ is an {\em
out-neighbor} of $x$, and $x$ is an {\em in-neighbor} of $y$. We
also denote it by $x\dom y.$ For disjoint sets $X,Y\subseteq V(H)$,
$X\dom Y$ means that $x\dom y$ for each $x\in X$ and $y\in Y$.

An {\em extension} of a digraph $G$ is a digraph $D$ obtained from
$G$ by replacing each vertex $u$ of $G$ by a set of independent
vertices $u_1,u_2,\ldots , u_{n(u)}$ such that for a pair $u,v$ of
vertices in $G$, $u_i\dom v_j$ in $D$ if and only if $u\dom v$ in
$G$.

For a bipartite digraph $H=(V,U;A)$, where $V$ and $U$ are its
partite sets, $H^{\dom}$ is the subdigraph induced by all arcs
directed from $V$ to $U$, $H^{\ldom}$ is the subdigraph induced by
all arcs directed from $U$ to $V$,  and $H^{\edom}$ is the
subdigraph induced by all 2-cycles of $H$, i.e., by the set $\{xy:\
xy\in A, yx\in A\}.$ The {\em converse} of $H$ is the digraph
obtained from $H$ by replacing every arc $xy$ with the arc $yx.$

We denote a directed cycle with $p$ vertices by $\vec{C}_p.$ For a
set $X$ of vertices of a digraph $H$, $D[X]$ denotes the subdigraph
of $H$ induced by $X$. For a digraph $H$, $UN(H)$ denotes the {\em
underlying graph} of $H$, i.e., an undirected graph obtained from
$H$ by disregarding all orientations and deleting multiple edges.

A digraph $D$ is {\em strong} (or, {\em strongly connected}) if
there is a directed path from $x$ to $y$ and a directed path from
$y$ to $x$ for every pair $x,y$ of vertices of $D$.

\noindent{\bf Forbidden family.} Let us introduced five special
digraphs for which, as we will see later, the minimum homomorphism
problem is NP-hard. The digraph $C_4'$ has vertex set
$\{x_1,x_2,y_1,y_2\}$ and arc set
$\{x_1y_1,y_1x_2,x_2y_2,y_2x_1,y_1x_1\}$. The digraph $C_4''$ has
the same vertex set, but its arc set is $A(C_4')\cup \{x_2y_1 \}$.
The digraph $H^*$ has vertex set $\{x_1,x_2,y_1,y_2,y_3\}$ and arc
set $$\{x_1y_1,y_1x_2,x_2y_2,y_2x_1,x_1y_3,x_2y_3 \}.$$

Let $N_1$ be a digraph with $V(N_1)= \{x_1,x_2,x_3,y_1,y_2,y_3\}$
and
$$A(N_1)=\{x_1y_1,y_1x_1, x_2y_2,y_2x_2, x_3y_3, y_3x_3, y_1x_2, y_1x_3,x_1y_2,x_1y_3,x_3y_2,x_2y_3\}.$$
Let $N_2$ be a digraph with $V(N_2)= \{x_1,x_2,x_3,y_1,y_2,y_3\}$
and
$$A(N_2)=\{x_1y_1, x_2y_2,y_2x_2, x_3y_3, y_3x_3, y_1x_2, y_1x_3,x_1y_2,x_1y_3,x_3y_2,x_2y_3\}.$$

A digraph $H$ belongs to the family $\cal HFORB$ if $H$ or its
converse is isomorphic to one of the five digraphs above or
$UN(H^s)$ is isomorphic to bipartite claw, bipartite net, bipartite
tent or even cycle with at least 6 vertices, where $s\in
\{\dom,\ldom,\edom\}.$

\noindent{\bf $k$-Min-Max ordering.} A collection $V_1,V_2,\ldots
V_k$  of subsets of a set $V$ is called a $k$-{\em partition} of $V$
if $V=V_1\cup V_2\cup \cdots \cup V_k$, $V_i\cap V_j=\emptyset$
provided $i\neq j$.
\begin{definition}\label{def}
Let $H=(V,A)$ be a digraph and let $k\ge 2$ be an integer. We say
that $H$ has a $k$-{\em Min-Max ordering} if there is a
$k$-partition of $V$ into subsets $V_1,V_2,\ldots V_k$  and there is
an ordering $v_1^i,v^i_2,\ldots, v^i_{\ell(i)}$ of $V_i$ for each
$i$ such that
\begin{description}
\item[(i)] Every arc of $H$ is an $(V_i,V_{i+1})$-arc for some $i \in \{1,2,\ldots ,k \}$,

\item[(ii)] $v_1^i,v^i_2,\ldots,
v^i_{\ell(i)}v^{i+1}_1v^{i+1}_2,\ldots, v^{i+1}_{\ell(i+1)}$ is a
Min-Max ordering of the subdigraph $H[V_i\cup V_{i+1}]$ for all $i
\in \{1,2,\ldots ,k \}$,
\end{description}
where all indices $i+1$ are taken modulo $k.$
\end{definition}


Note that if $H$ is a strong digraph in which the greatest common
divisor of all cycle lengths is $k$, then $V(H)$ has a
$k$-partition, $k\ge 2$, satisfying (i) (see Theorem 10.5.1 in
\cite{bang2000}). A simple example of a digraph having a $k$-Min-Max
ordering is an extension of $\vec{C}_k$.

\noindent{\bf Dichotomy and paper organization.} The main result of
this paper is the following:

\begin{theorem}\label{mainth}
Let $H$ be an semicomplete bipartite digraph. If $H$ contains a
digraph from $\cal HFORB$ as an induced subdigraph, then MinHOM($H$)
is NP-hard.

Assume that P$\neq$NP. Then the following three assertions are
equivalent:

(i) MinHOM($H$) is polynomial time solvable;

(ii) $H$ does not contain a digraph from $\cal HFORB$ as an induced
subdigraph;

(iii) Each component of $H$ has a $k$-Min-Max ordering for $k=2$ or
4.
\end{theorem}

Theorem \ref{mainth} follows from Corollaries \ref{corP} and
\ref{corNP}.

The rest of the paper is organized as follows. In Section
\ref{ksec}, we study properties of $k$-Min-Max orderings. In Section
\ref{psec}, we prove polynomial cases of MinHOM($H$) when $H$ is a
semicomplete bipartite digraph. In Section \ref{npsec}, we establish
NP-hard cases of the problem. In Section \ref{gdsec} we formulate a
dichotomy for the computational complexity of MinHOM($H$) when $H$
is a semicomplete multipartite digraph. Section \ref{frsec} provides
a short discussion of further research.

\section{Properties of $k$-Min-Max Orderings}\label{ksec}

Digraphs having $k$-Min-Max ordering have a very special structure
as described in the following lemma.

\begin{lemma}\label{monL}
If a strong digraph $H=(V,A)$ has a $k$-Min-Max ordering as
described in Definition \ref{def}, then we can define $0 \leq
L(i,j) < R(i,j) \leq \ell(j)+1$ for all $i$ and $j$ such that
\begin{description}
  \item[(a):] $N^+(v_i^j)=\{v_{L(i,j)+1}^{j+1},v_{L(i,j)+2}^{j+1},
\ldots ,v_{R(i,j)-1}^{j+1}\}$,  where all superscripts are taken
modulo $k$;
  \item[(b):] For all $j$ and $i<i'$ we have $R(i,j) \leq R(i',j)$ and $L(i,j) \leq L(i',j)$.
\end{description}
\end{lemma}
\pf Suppose that $v^j_iv^{j+1}_{m-1},v^j_iv^{j+1}_{m+1}\in A$, but
$v^j_iv^{j+1}_m\not\in A$. Since $H$ is strong, there is an arc
$v^j_tv^{j+1}_m$ in $H$. By the definition of a $k$-Min-Max
ordering, $v^j_iv^{j+1}_m\in A$, a contradiction that proves (a).
Similarly, one can show (b). \qed

\2

The construction used in the following theorem was inspired by
somewhat similar constructions in \cite{khannaSIAMJC30} and
\cite{cohenJAIR22}.

\begin{theorem}\label{mm2}
If a digraph $H$ has a $k$-Min-Max ordering, then MinHOM$(H)$ is
polynomial-time solvable.
\end{theorem}
\pf Let $V_1, V_2,\ldots , V_k$ be defined as in Definition
\ref{def}. Let $D$ be an input digraph. Assume that there is
homomorphism $f$ of $D$ to $H.$ Let $G_1, G_2, \ldots, G_k$ be a
$k$-partition of $V(D)$ such that  $f(G_j)\subseteq V_{j}$ for each
$j \in \{1,2,\ldots,k\}$. Observe that all arcs in $D$ are
$(G_j,G_{j+1})$-arcs, where all indices are taken modulo $k$ and $j
\in \{1,2,3,\ldots ,k\}$.

We will now show how to find a minimum cost homomorphism of $D$ to
$H$, where the vertices of $G_j$ are mapped to $V_j$ for all
$j=1,2,\ldots,k$. We will build a directed graph $\cal L$ with
vertex set $\cup_{j=1}^k(G_j \times V_j)$ together with two other
vertices, denoted by $s$ and $t$. We will also denote $t$ by
$(x,v_{\ell(j)+1}^j)$ for every $j \in \{1,2,\ldots,k\}$. The
weighted arcs of $\cal L$ are as follows, where $M$ is any constant
greater than the cost of a minimum cost homomorphism of $D$ to $H$.

\begin{itemize}
\item An arc from $s$ to $(x,v_1^j)$, of weight $\infty$, for each
$x \in G_j$.

\item An arc from $(x,v_i^j)$ to $(x,v_{i+1}^j)$, of weight
$c_i(x)+M$, for each $x \in G_j$ and $i \in
\{1,2,\ldots,\ell(j)\}$. Recall that when $i=\ell(j)$ the arc
enters $t$.

\item an arc from $(x,v_i^j)$ to $(y,v_{L(i,j)+1}^{j+1})$ and an
arc from $(y,v_{R(i,j)}^{j+1})$ to $(x,v_{i+1}^j)$ for every $xy
\in A(D)$ with $x \in G_j$ and every $i=1,2,\ldots,\ell(j)$.
Furthermore the weight of these arcs are $\infty$.
\end{itemize}

\vspace{2mm} A {\em cut} in $\cal L$ is a partition of the
vertices into two sets $S$ and $T$ such that $s \in S$ and $t \in
T$ and the weight of a cut is the sum of weights of all arcs going
from a vertex of $S$ to a vertex of $T$. We will show that the
minimum weight cut in $\cal L$ has weight equal to the minimum
cost homomorphism of $D$ to $H$ plus $|V(D)| M$.

Let $f$ be a minimum cost homomorphism of $D$ to $H$, and assume
that $f(x)=v_{a(x)}^{j}$ for each $x \in G_j$ and for all
$j=1,2,\ldots,k$. Define a cut in $D$ as follows: $S=\{
(x,v_i^j):\ i \leq a(x), \mbox{ } j=1,2,\ldots,k \} \cup \{s\}$
and $T=V(L)-S$. Note that the arcs from $(x,v_{a(x)}^j)$ to
$(x,v_{a(x)+1}^j)$ belong to the cut and contribute
$c_{f(x)}(x)+M$ to the weight of the cut. We will now show that
there are no arcs of infinite weight in the cut, which would imply
that the weight of $S$ is exactly the cost of a minimum cost
homomorphism from $D$ to $H$ plus $|V(D)| M$.

Clearly no arc out of $s$ belongs to the cut $S$. Assume for the
sake of contradiction that the arc $(x,v_i^j)$ to
$(y,v_{L(i,j)+1}^{j+1})$ belongs to the cut $S$ for some $xy \in
A(D)$ with $x \in G_j$. This implies that $a(x) \geq i$ (as
$(x,v_i^j) \in S$) and $ a(y) <  L(i,j)+1$ (as
$(y,v_{L(i,j)+1}^{j+1}) \not\in S$). By Lemma \ref{monL} (b), this
implies that $a(y) \leq L(i,j) \leq L(a(x),j)$. Thus, there is no
arc from $v_{a(x)}^j$ to $v_{a(y)}^{j+1}$ in $H$, by the
definition of $L(i,j)$. This is a contradiction to $f$ being a
homomorphism.

Now assume for the sake of contradiction that the arc
$(y,v_{R(i,j)}^{j+1})$ to $(x,v_{i+1}^j)$ belongs to the cut $S$
for some $xy \in A(D)$ with $x \in G_j$. This implies that $a(x) <
i+1$ (as $(x,v_{i+1}^j) \not\in S$) and $ a(y) \geq  R(i,j)$ (as
$(y,v_{R(i,j)}^{j+1}) \in S$). By Lemma \ref{monL} (b), this
implies that $a(y) \geq R(i,j) \geq R(a(x),j)$. Thus, there is no
arc from $v_{a(x)}^j$ to $v_{a(y)}^{j+1}$ in $H$, by the
definition of $R(i,j)$. This is a contradiction to $f$ being a
homomorphism. We have now proved that the cut $S$ has the stated
weight.

For the sake of contradiction assume that there exists a cut,
$S'$, in $\cal L$ of smaller weight than $S$. As the weight of $S$
is less than $M+|V(D)|M$ we note that the cut $S'$ contains
exactly one arc of the form $(x,v_i^j) (x,v_{i+1}^j)$ for each $x
\in V(D)$. Therefore we may define a mapping, $f'$, from $V(D)$ to
$V(H)$ by letting $f'(x)=v_i^j$ if and only if $(x,v_i^j) \in S'$
and $(x,v_{i+1}^j) \not\in S'$. We will now show that $f'$ is a
homomorphism of $D$ to $H$ of smaller cost than $f$, a
contradiction. This would imply that $S$ is a minimum weight cut,
and we would be done.

Note that if $f'$ is a homomorphism of $D$ to $H$, then it has
smaller cost than $f$, as $S'$ is a cut of smaller weight than
$S$. Let $xy$ be any arc in $D$, and assume without loss of
generality that $x \in G_j$. Let $i_x$ and $i_y$ be defined such
that $(x,v_{i_x}^j) \in S'$ and $(x,v_{i_x+1}^j) \not\in S'$
(i.e., $f'(x)=v_{i_x}^j$) and $(y,v_{i_y}^{j+1}) \in S'$ and
$(y,v_{i_y+1}^{j+1}) \not\in S'$ (i.e., $f'(y)=v_{i_y}^{j+1}$). As
the arc $(x,v_{i_x}^j)(y,v_{L(i_x,j)+1}^{j+1})$ is not in the cut,
we must have $(y,v_{L(i_x,j)+1}^{j+1}) \in S$, which implies that
$i_y \geq L(i_x,j)+1$.  Furthermore as the arc
$(y,v_{R(i_x,j)}^{j+1})(x,v_{i_x+1}^j)$ is not in the cut, we must
have $(y,v_{R(i_x,j)}^{j+1}) \not\in S$, which implies that $i_y <
R(i_x,j)$. We have now shown that $L(i_x,j)+1 \leq i_y <
R(i_x,j)$, which by the definition of the functions $L$ and $R$
implies that $v_{i_x}^j v_{i_y}^{j+1}$ is an arc in $H$. Therefore
$f'$ is a homomorphism. \qed

\section{Polynomial Cases}\label{psec}

We start from a special case which is of importance when $H$
contains no induced $\vec{C}_4.$

\begin{lemma} \label{monoton}
Let $H=(V,U;A)$ be a semicomplete bipartite digraph, which does
not contain an induced subdigraph belonging to $\cal HFORB$ or an
induced directed $4$-cycle. Suppose for every $v,v' \in V$ we have
$N^+(v) \subseteq N^+(v')$ or $N^+(v') \subseteq N^+(v)$. Then $H$
has a 2-Min-Max ordering.
\end{lemma}
\pf We say that vertices $v_i, v_j \in V$ are {\em similar} if
$N^+(v_i)= N^+(v_j)$. Consider similarity classes $V_1,V_2,\ldots
,V_s$ of $V$. Moreover assume that $N^+(V_i) \subset N^+(V_j)$ for
$i < j$. Set $U_1= N^+(V_1)$ and $U_i= N^+(V_i)-\cup_{j=1}^{i-1}
U_j$ for each $i>1$. If $s=1$ then $UN(H^{\dom})$ is a complete
bipartite graph and a Min-Max ordering of $H^{\ldom}$ (which
exists by Theorem \ref{unddich}) is a 2-Min-Max ordering of $H$.
Assume $s>1$. We prove the following two claims:

\begin{description}
\item[(1)] Let $u_i\in U_i,\ v_j\in V_j$, $j>i$, and $u_i v_j \in
A$. Then $U_r \dom v_j$ for each $r > i$ and $u_i \dom V_t$ for
each
$t <j$.\\

{\em Proof of (1):} By the definition of $U_i$ and $V_i$, if $r >
j$ then $U_r \dom v_j$. Now suppose that $i<r \leq j$ and $u_r v_j
\not\in A$ for some $u_r\in U_r.$ Let $v_i \in V_i$ be arbitrary,
and note that $H[\{u_i,v_i,u_r,v_j\}]$ is isomorphic to $C_4'$ or
$C_4''$, a contradiction. So $u_rv_j \in A$.

 By the definition of $U_i$ and $V_i$, if $t<i$
then $u_i \dom V_t$.  Now suppose that $i \leq t < j$ and that
$u_iv_t \not\in A$, for some $v_t \in V_t$.
 Let $u_j \in U_j$ be arbitrary, and note
that $H[\{u_i,v_t,u_j,v_j\}]$ is isomorphic to $C_4'$ or $C_4''$,
a contradiction.

\item[(2)] If $u,u' \in U$ then $N^+(u) \subseteq N^+(u')$ or
$N^+(u') \subseteq N^+(u)$. \\

{\em Proof of (2):} Assume that this is not the case, and there
exist $v \in N^+(u)-N^+(u')$ and $v' \in N^+(u')-N^+(u)$. This
implies that $u \not= u'$, $v \not=v'$, $uv,u'v' \in A$ and
$uv',u'v \not\in A$.
  If $u \in U_i$ and $u' \in U_j$ and
$i<j$, then by (1) we note that $u' v \in A$ a contradiction. So
for some $i$ we must have $\{u,u'\} \subseteq U_i$.
  Analogously, if  $v \in V_a$ and $v' \in V_b$ and
$a<b$, then by (1) we note that $u' v \in A$, a contradiction. So
for some $j$ we must have $\{v,v'\} \subseteq V_j$. If $i>j$ then
$U_i \dom V_j$, which is a contradiction, so we must have $i \leq
j$.

If $i<j$ then let $u'' \in U_j$ and $v'' \in V_i$ be arbitrary.
Note that by (1) we must have the arcs $u''v,u''v',uv'',u'v''$ in
$H$. By the construction of the sets $U_i,U_j,V_i,V_j$ we now note
that the underlying graph of  $H[\{u,u',u'',v,v',v''\}]^{\edom}$
is the $6$-cycle $v'' u v u'' v' u' v''$, a contradiction.

Therefore we must have $i=j$. First assume that $i<s$. Now
consider $v_s \in V_s$ and $u_s \in U_s$. By (1) there is no arc
from $\{u,u'\}$ to $v_s$. However $H[\{u,u',v,v',v_s,u_s\}]$ is
either $N_1$ or $N_2$, a contradiction.

Similarly for the case $i=s$ we derive a contradiction.
\end{description}

Now consider an ordering $(u_1,u_2,\ldots ,u_a)$ of the vertices
in $U$ and an ordering $(v_1,v_2,\ldots ,v_b)$ of the vertices in
$V$, defined as follows. If $i<j$ then $d^+(v_i) \leq d^+(v_j)$
and if $d^+(v_i) = d^+(v_j)$ then $d^-(v_i) \geq d^-(v_j)$.
Furthermore when  $i<j$ then $d^+(u_i) \leq d^+(u_j)$ and if
$d^+(u_i) = d^+(u_j)$ then $d^-(u_i) \geq d^-(u_j)$. Note that the
ordering $(v_1,v_2,\ldots ,v_b)$ first contains vertices from
$V_1$ then from $V_2$, etc.

We will now show that for every $i \in \{1,2,\ldots ,b\}$ there
exists an integer $\alpha_i$ such that $N^+(u_i) = \{v_1,v_2,
\ldots ,v_{\alpha_i} \}$. Suppose this is not the case. Then there
exists an arc $u_i v_j$ in $H$, such that $ u_i v_{j-1}$ is not an
arc in $H$. Thus, both $v_j$ and $v_{j-1}$ belong to some $V_k$,
as otherwise we have a contradiction to (1). This implies that
$d^+(v_j)=d^+(v_{j-1})$ and $d^-(v_{j-1}) \geq d^-(v_j)$. Note
that every vertex $u \in N^-(v_{j-1})$ has $N^+(u_i) \subseteq
N^+(u)$, by (2). Therefore $u\dom v_j$. However this implies that
$d^-(v_{j-1}) < d^-(v_j)$ (as $u_i \dom v_j$ but $u_i$ does not
dominate $v_{j-1}$), a contradiction.

Using the fact that $N^+(u_i) = \{v_1,v_2, \ldots ,v_{\alpha_i} \}$
for each $i \in \{1,2,\ldots ,b\}$ and that $\alpha_i\ge
\alpha_{i-1}$ for each $i \in \{2,3,\ldots ,b\}$ (as $d^+(u_i) \geq
d^+(u_{i-1})$) and the similar relations for the vertices of $V$, we
can readily conclude that $H$ has a 2-Min-Max ordering.\qed

\2

The {\em distance} $\dist(x,y)$ between a pair $x,y$ of vertices
in an undirected graph $G$ is the length of the shortest path
between $x$ and $y.$ The {\em diameter} of $G$ is the maximal
distance between a pair of vertices in $G.$

\2

The following theorem shows when MinHOM($H$) is polynomial time
solvable if $H$ is strong and does not contain $\vec{C}_4$ as an
induced subdigraph.

\begin{theorem} \label{mm2exist}
Let $H$ be a strongly connected semicomplete bipartite digraph.
Assume that $H$ does not contain a digraph from $\cal HFORB$ or
$\vec{C}_4$ as an induced subdigraph. Then $H$ has a 2-Min-Max
ordering and MinHOM($H$) is polynomial time solvable.

Furthermore, either $UN(H^{\dom})$ or $UN(H^{\ldom})$ are complete
bipartite graphs, or the following holds.
   For every pair $u,u'$ of distinct
vertices of $U$ we have $N^+(u)\subseteq N^+(u')$ or
$N^+(u')\subseteq N^+(u)$ and for every pair $v,v'$ of distinct
vertices of $V$ we have $N^+(v)\subseteq N^+(v')$ or
$N^+(v')\subseteq N^+(v)$.

\end{theorem}
\pf By Theorem \ref{mm2}, to prove the first part part of this
theorem (before `Furthermore'), it suffices to show that $H$ has a
2-Min-Max ordering. Let $V$ and $U$ be partite sets of $H$. Denote
$H_1=H^{\dom}$ and $H_2=H^{\ldom}.$ It follows from Theorem
\ref{unddich} that $UN(H_1)$ and $UN(H_2)$ have Min-Max orderings
and so do $H_1$ and $H_2.$

Let $d_i$ be the diameter of $UN(H_i)$, $i=1,2.$ Observe that if
$UN(H_1)$ ($UN(H_2)$) is a complete bipartite graph, then a
Min-Max ordering of $H_2$ ($H_1$) is a $2$-Min-Max ordering of
$H$. Therefore, MinHOM($H$) is polynomial time solvable by Theorem
\ref{mm2}. Notice that $UN(H_i)$ is complete bipartite if and only
if $d_i=2$. Thus, we may assume that both $d_1\ge 3$ and $d_2\ge
3.$

We consider the following cases for the value of $d_1\ge 3$.

\2

\noindent{\bf Case 1:} $d_1> 4$.

We will show that $UN(H_2)$ is a complete bipartite graph or,
equivalently, $d_2=2.$ Assume that $d_1$ is odd, as the case of
$d_1$ even can be considered similarly. Let $P=v_1u_1v_2u_2\ldots
v_{k-1}u_{k-1}v_ku_k$ be a shortest path of length $d_1$ between
$v_1$ and $u_k$ in $UN(H_1)$. Let $\hat{U}=\{u_1,u_2,\ldots ,u_k\}$
and $\hat{V}=\{v_1,v_2,\ldots ,v_k\}$. We will first prove that
$\hat{U}\dom \hat{V}.$ Since $P$ is a shortest path, we have
$u_i\dom v_j$ for each $u_i\in \hat{U}$ and $v_j\in \hat{V}$
provided $j\not\in \{i,i+1\}.$ Thus, it sufficient to prove
 \begin{equation}\label{*}
u_i\dom v_i\ (i=1,2,\ldots ,k) \mbox{ and } u_i\dom v_{i+1}\
(i=1,2,\ldots ,k-1)
 \end{equation}

Consider the subdigraph $H'$ of $H$ induced by four vertices
$v_i,u_i,v_{i+2},u_{i+1}$, where $i\in \{1,2,\ldots, k-2\}.$ By the
definition of $P$ (including the fact that $P$ is a shortest path),
we have $v_iu_i,v_{i+2}u_{i+1},u_{i+1}v_i,u_iv_{i+2}\in A(H)$, but
$v_iu_{i+1},v_{i+2}u_i\not\in A(H)$. Since $H'$ is not isomorphic to
either $\vec{C}_4$ or $C_4'$, we have $u_iv_i, u_{i+1}v_{i+2} \in
A(H)$. This proves that $u_i\dom v_i$ provided $i=1,2,\ldots ,k-2$
and $u_i\dom v_{i+1}$ provided $i=2,3,\ldots ,k-1.$ Thus, to prove
(\ref{*}) it remains to show  that
 \begin{equation}\label{**}
u_{k-1}\dom v_{k-1},\ u_k\dom v_k,\ u_1\dom v_2
 \end{equation}

Consider the subdigraph $H''$ of $H$ induced by four vertices
$v_{k-1},u_{k-1},v_{k},u_{k}.$ By the definition of $P$, we have
$v_{k-1}u_{k-1},v_ku_k,v_ku_{k-1}, u_kv_{k-1}\in A(H)$, but
$v_{k-1}u_k\not\in A(H)$. We have proved that $u_{k-1}\dom v_k$.
Since $H''$ is not isomorphic to $C'_4$ or $C_4''$, we have
$u_{k-1}v_{k-1},u_kv_k\in A(H)$.

Consider $H[\{v_2,v_3,u_1,u_2\}]$. By the definition of $P$, we
have $v_2u_1,v_2u_2,v_3u_2,u_1v_3\in A(H)$, but $v_3u_1\not\in
A(H)$. We have proved that $u_2\dom v_2$. Since
$H[\{v_2,v_3,u_1,u_2\}]$ is not isomorphic to $C'_4$ or $C_4''$,
we have $u_1v_2,u_2v_3\in A(H)$. This implies that (\ref{**}) and,
thus, (\ref{*}) has been proved.

Consider vertex $u \in U- \hat{U}$; we will show that $u
\rightarrow \{v_1,v_2,\ldots ,v_k\}$. Suppose this is not true.
Let $j$ be the smallest index such that $uv_j \not\in A(H)$. We
have $v_ju \in A(H)$. Suppose $j>1$. Since $H[\{u,v_1,u_1,v_j\}]$
is not isomorphic to $C_4'$ or $C_4''$, we have $v_ju_1,v_1u \in
A(H)$. Since $P$ is a shortest path, we have $j=2$ as otherwise
$v_1uv_ju_{j+1}\ldots v_ju_k$ is shorter than $P$. We have $v_3u
\not\in A(H)$ as otherwise $v_1 u v_3 u_3\ldots v_ku_k$ is shorter
than $P$. However, $C_4'$ is isomorphic to $H[\{v_2,u,v_3,u_3\}]$,
a contradiction.

Now assume that $j=1$. We have $v_3u \not\in A(H)$ as otherwise we
have a shorter path. Since $H$ is semicomplete bipartite, we have
$uv_3 \in A(H)$. However $H[\{v_1,u,v_3,u_2\}]\cong C_4'$, a
contradiction.

Analogously we can prove that $\hat{U} \rightarrow v$ for every
$v\in V-\hat{V}$. Consider $u \in U-\hat{U},\ v \in V-\hat{V}$. We
show that $uv \in A(H)$. Suppose this is not true. We have $vu \in
A(H)$. Since $H[\{v,u,v_1,u_1\}]$ is not isomorphic to $\vec{C}_4,\
C_4'$ or $C_4''$ we have $v_1u,vu_1 \in A(H)$. Since
$H[\{v,u,v_{k},u_k \}]$ is not isomorphic to $C_4'$ or $C_4''$ we
have $v_ku,vu_k \in A(H)$. But now ${\rm dist}(v_1,u_k) =3$ in
$UN(H_1)$, a contradiction.

\2

\noindent{\bf Case 2:} $d_1=4$.

We will show that again $UN(H_2)$ is a complete bipartite graph.
Assume that $v_1u'_1v'_2u'_2v'_3$ is a shortest path between a pair
$v_1\in V$ and $v'_3\in U$ in $UN(H_1).$ Let $U_1=N^+(v_1)$, $V_2=
N^-(U_1)-\{v_1\}$, $U_2=N^+(V_2)-U_1$ and $V_3=N^-(U_2)-V_2$. By the
definitions, $V=\{v_1\}\cup V_2\cup V_3$ and $U=U_1\cup U_2.$
Observe also that $(U_1,V_3)=U_1\times V_3$, $(V_3,U_1)=\emptyset$,
$(U_2,\{v_1\})=U_2\times \{v_1\}$ and $(\{v_1\},U_2)=\emptyset.$ Let
$u_1\in U_1$, $u_2 \in U_2$, $v_2\in V_2$ and $v_3 \in V_3$ be
arbitrary. If $u_2v_3 \not\in A(H)$, then $v_3u_2 \in A(H)$, and now
$H[\{u_1,v_3,u_2,v_1 \}]$ is either $\vec{C}_4$ or $C_4'$, a
contradiction. Therefore, we have $u_2v_3 \in A(H)$ and consequently
$(U_2,V_3)=U_2\times V_3$. Consider $v_3, u_2$ where $v_3u_2 \in
A(H)$. Since $H[\{u_1,v_3,u_2,v_1 \}]$ is not $C_4'$ or $\vec{C}_4$,
we conclude that $u_1v_1 \in A(H)$ and consequently
$(U_1,\{v_1\})=U_1\times \{v_1\}$.


Note that we have already proved that the underlying graph of $H_2[
U_1 \cup U_2 \cup V_3 \cup \{v_1\}]$ is a complete bipartite graph.
Suppose $u_1v_2 \in A(H)$. Then $u_2v_2 \in A(H)$ as otherwise
$H[\{u_1,v_1,u_2,v_2\}]$ is isomorphic to $C_4'$ or $C_4''$, a
contradiction. Therefore, if $u_1v_2 \in A(H)$, then
$(U_2,\{v_2\})=U_2\times \{v_2\}$. Thus, to show that $(U_2,
V_2)=U_2\times V_2$ it suffices to prove that every vertex in $V_2$
has an in-neighbor in $U_1$. Suppose this is not true, and let $X_2$
be the set of all vertices in $V_2$ that does not have an
in-neighbor in $U_1$.  Let $Y_2$ be the set of all vertices in $U_2$
that have an arc into $X_2$. As $H$ is strong some vertex in $H$
must have an arc into $X_2$, which implies that $Y_2 \not=
\emptyset$.

If there is an arc $v_3 y_2$ from $V_3$ to $Y_2$, then let $x_2$
be an out-neighbor of $y_2$ in $X_2$ and let $u_1 \in U_1$ be
arbitrary. However this is a contradiction to
$H[\{v_3,y_2,x_2,u_1\}]$ not being isomorphic to $C_4'$ and
$C_4''$, which implies that there is no arc from $V_3$ to $Y_2$.

As $H$ is strong and there is no arc from $V_3$ into $U_1$ or
$Y_2$, there must be an arc, say $v_3u_2$, from $V_3$ into
$U_2-Y_2$. As $H$ is strong there must also be an arc from
$V(H)-X_2-Y_2$ into $X_2 \cup Y_2$. By the above this arc, say
$v_2 y_2$, must be from $V_2-X_2$ to $Y_2$. As $y_2$ belongs to
$Y_2$ there must be a vertex, say $x_2 \in X_2$, such that $y_2
x_2 \in A(H)$. As $v_2 \not\in X_2$ we note that there is a
vertex, say $u_1 \in U_1$, such that $u_1 v_2 \in A(H)$. As
$u_1x_2 \not\in A(H)$ and  $H[\{v_2,x_2,u_1,y_2\}]$ is not
isomorphic to $C_4'$ and $C_4''$, we note that $x_2y_2, v_2u_1 \in
A(H)$. if $u_2v_2 \not\in A(H)$, then $H[\{v_1,v_2,u_1,u_2\}]$ is
isomorphic to $C_4'$ and $C_4''$ (as $v_1u_2 \not\in A(H)$), a
contradiction. As $u_2v_2 \in A(H)$ and $H[\{v_2,v_3,u_2,y_2\}]$
is not isomorphic to $C_4'$ and $C_4''$ we note that $y_2 v_2 \in
A(H)$ (as $v_3y_2 \not\in A(H)$). As $H[\{v_2,x_2,u_2,y_2\}]$ is
not isomorphic to $C_4'$ and $C_4''$ we note that $v_2 u_2 \in
A(H)$ (as $u_2x_2 \not\in A(H)$). However, the underlying graph of
$H[v_1,u_1,v_2,v_3,u''_2,v''_2,u_2]^{\edom}$ is now a bipartite
claw, with edges $\{v_2u_1, v_2y_2, v_2u_2, u_1v_1, y_2x_2,
u_2v_3\}$, a contradiction. Therefore $U_2 \dom V_2$.

We show that $u_1v_2 \in A(H)$ for every $u_1 \in U_1$ and $v_2
\in V_2$. Suppose this is not true for some $v_2\in V_2$ and
$u_1\in U_1$. Then we have $v_2u_1 \in A(H)$. Let $v_3\in V_3$ be
arbitrary and $u_2\in U_2\cap N^+(v_3).$ Then
$H[\{u_1,v_3,u_2,v_2\}]$ is isomorphic to $C_4'$ or $C_4''$. This
completes our proof that $UN(H_2)$ is a complete bipartite graph.

\2

\noindent{\bf Case 3:} $d_1=3$.

Consider a Min-Max ordering $\pi$ for $H_1$. Let
$\pi(x)=\min\{\pi(x'):\ x'\in V\}$, $\pi(t)=\max\{\pi(t'):\ t'\in
U\},$ $\pi(y)=\max\{\pi(y'):\ y'\in U,\ x\dom y'\},$ and
$\pi(z)=\min\{\pi(z'):\ z'\in V,\ z'\dom t\}.$

Let $T=N^-(t).$ Since $H$ is strong, every vertex of $V$ has an
out-neighbor. Since $\pi$ is a Min-Max ordering, $z'\dom t$ for
each $z'\in V$ with $\pi(z')\ge \pi(z)$. Thus, $T=\{z'\in V:\
\pi(z')\ge \pi(z)\}.$ Let $X=N^+(x).$ Since $H$ is strong, every
vertex of $U$ has an in-neighbor. Since $\pi$ is a Min-Max
ordering, $x\dom y'$ for each $y'\in U$ with $\pi(y')\le \pi(y).$
Thus, $X=\{y'\in U:\ \pi(y')\le \pi(y)\}.$

Since $\dist(x,z)=2$ in $UN(H_1)$, we have $z\dom y'$ for some
$y'\in N^+(x)$. Since $\pi$ is a Min-Max ordering, $z\dom y$
(consider the arcs $zy'$ and $xy$). Now for every $z''\in V$ with
$\pi(z'')\le \pi(z)$, we have $z''\dom y$ (as $\pi$ is a Min-Max
ordering). Similarly, for every $y''\in U$ with $\pi(y'')\ge
\pi(y)$, we have $z\dom y''.$

Let $\pi(w)=\min \{\pi(w'):\ w'\in U\}.$ Notice that
$\dist(w,t)=2$ in $UN(H_1)$. Thus, for some $z'\in T$, we have
$z'\dom w$. Hence, $z\dom w$ and $z\dom X$ (as $\pi$ is a Min-Max
ordering). We conclude that $z\dom U.$ Similarly, we can obtain
that $V\dom y.$

Let $Y=V-T$ and $Z=U-X.$ Let $x'\in X,\ y'\in Y.$ We have $z\dom
x'$ and $y'\dom y$. Hence (as $\pi$ is a Min-Max ordering),
$y'\dom x'.$ Thus, $Y\dom X.$ Analogously, we can prove that
$T\dom Z.$

Let $x'\in X$, $t'\in T$, $y'\in Y$ and $x'\dom t'$. Then $t't,\
ty',y'x'\in A(H)$. Since $H[\{x',t,t',y'\}]$ is not isomorphic to
$\vec{C}_4$, $C_4'$ or $C_4''$, we have $x'\dom y'$ and $t\dom
t'.$ Thus, if $x'\dom t'$, we have $x'\dom Y$. Analogously, if
$x'\dom t'$, we have $Z\dom t'.$ Hence, \begin{equation}\label{ee}
x'\dom t' \mbox{ implies } x'\dom Y \mbox{ and } Z\dom t'
\end{equation}

We will now prove the following: for a pair $u,u'$ of distinct
vertices of $U$ we have $N^+(u)\subseteq N^+(u')$ or
$N^+(u')\subseteq N^+(u)$. By Lemma \ref{monoton}, this implies
that $H$ has a 2-Min-Max ordering and we are done. Suppose that we
have neither $N^+(u)\subseteq N^+(u')$ nor $N^+(u')\subseteq
N^+(u)$. Thus, there is a pair $v,v'$ of vertices in $V$ such that
$u\dom v$, $u'\dom v'$, but  $uv'$ and $u'v$ are not arcs in $H$.
Since $H[\{u,u',v,v'\}]$ is not isomorphic to $\vec{C}_4$, $C_4'$
and $C_4''$, we have $v\dom u$ and $v'\dom u'.$ Now consider four
cases.

\2

\noindent{\bf Case 3.1:} $v,v'\in Y.$ Let $t'\in T$. By the
definition of $t$, $t\not\in \{u,u'\}.$ If $u'\dom t'$, then
$H[\{t,t',v,u'\}]$ is isomorphic to $\vec{C}_4$, $C_4'$ or $C_4''$,
which is impossible. Thus, $u't'\not\in A(H)$ and $t'\dom u'.$
Analogously, $ut'\not\in A(H)$ and $t'\dom u.$ By the fact that
$t'\dom t$ and the existence and nonexistence of previously
considered arcs, we conclude that $H[\{v,v',u,u',t,t'\}]$ is
isomorphic to $N_1$ or $N_2,$ which is impossible.

\2

\noindent{\bf Case 3.2:} $u,u'\in Z$. We can show that this case is
impossible similarly to Case 3.1 but considering $x,w$ instead of
$t,t'.$

\2

\noindent{\bf Case 3.3:} $v\in Y$, $v'\in T$. By (\ref{ee}), $u'\in
Z$. By Case 3.2, we may assume that $u\in X.$ Then
$H[\{v,v',u',t\}]$ is isomorphic to $\vec{C}_4$, $C_4'$ or $C_4''$,
which is impossible.

\2

\noindent{\bf Case 3.4:} $v,v'\in T$. By Case 3.2, we may assume
that $u\in X.$ By (\ref{ee}), $Z\dom v$ and, thus, $u'\in X.$ By
(\ref{ee}), we conclude that $uxu,\ u'xu',\ vtv$ and $v'tv'$ are
2-cycles. Notice that $t\dom x$, but $xt\not\in A(H)$. Now it
follows that $H[\{x,v,v',u,u',t\}]^{\edom}$ is isomorphic to $C_6$,
a contradiction.

\2 \2

It follows from Cases 1,2 and 3 that if neither $UN(H_1)$ nor
$UN(H_2)$ are complete bipartite graphs, then we must have
$d_1=d_2=3$. In this case we have shown that for every pair $u,u'$
of distinct vertices of $U$ we have $N^+(u)\subseteq N^+(u')$ or
$N^+(u')\subseteq N^+(u)$. However, by swapping the roles of $U$ and
$V$ we also get that  for every pair $v,v'$ of distinct vertices of
$V$ we have $N^+(v)\subseteq N^+(v')$ or $N^+(v')\subseteq N^+(v)$.
\qed

The following theorem shows when MinHOM($H$) is polynomial time
solvable for the case when $H$ is not strong, and does not contain
$\vec{C}_4$ as an induced subdigraph.

\begin{theorem} \label{nonstrong}
Let $H=(U,V;A)$ be a semicomplete bipartite digraph with strong
components $C_1,C_2,\ldots ,C_p$ ($p\ge 2$) satisfying the
following:
\begin{itemize}
\item There is no arc from $C_i$ to $C_j$ for $i>j$, \item  $H$
does not contain an induced subdigraph belonging to $\cal HFORB$ or
an induced directed $4$-cycle.
\end{itemize}
Then $H$ has a 2-Min-Max ordering and MinHOM($H$) is polynomial time
solvable.
\end{theorem}
\pf Suppose there are $v,v' \in V$ and $u,u'\in U$ such that
$A(H^{\dom}[\{v,v',u,u'\}])=\{vu,v'u' \}$. Note that $\{v,v',u,u'\}$
belong to a strong component of $H$ as they are contained in a
$4$-cycle. Let $\{v,v',u,u'\} \subseteq V(C_t)$ for some $t$ and let
$V_1= \{ x \in V \ | \ x \in C_i \ , \ i < t\}$, $U_1=\{ y \in U \ |
\ y \in C_i \ , \ i < t\}$, $V_2= V \cap C_t$, $U_2= U \cap C_t$ ,
$V_3=V-V_1-V_2$ and $U_3=U-U_1-U_2$.

If there is a $w' \in U_3$ and $w \in V_3$, such that $w' \dom w$,
then $H[\{u,v,u',v', w,w'\}]$ is either the dual of $N_1$ or $N_2$.
Therefore we must have $A(H[U_3 \cup V_3])=V_3 \times U_3$.
Analogously we must have $A(H[U_1 \cup V_1])=V_1 \times U_1$.
Consider $H'=H[U_2 \cup V_2]$ and note that $H'$ is strong and does
not contain a digraph from $\cal HFORB$ or $\vec{C}_4$ as an induced
subdigraph. Therefore Theorem \ref{mm2exist} implies that $U_2 \dom
V_2$ (as $V_2 \dom U_2$ is not true), which furthermore implies that
$(U_1 \cup U_2) \dom (V_2 \cup V_3)$.

Let $\pi{}$ be a  min-max ordering of $H^{\dom}$. Let $uv \in A(H)$
and $u'v' \in A(H)$ be two distinct arcs from $U$ to $V$. As
$d^+(u)>0$ we note that $ u \in U_1 \cup U_2$, by the above.
Analogously we note that $u' \in U_1 \cup U_2$, $v \in V_2 \cup V_3$
(as $d^-(v)>0$) and $v' \in V_2 \cup V_3$. By the above we therefore
have $uv',u'v \in A(H)$.  As $u,u',v,v'$ were chosen arbitrarily,
this implies that $\pi{}$ is a  2-Min-Max ordering.

If there are no $v,v' \in V$ and $u,u'\in U$ such that
$A(H^{\dom}[\{v,v',u,u'\}])=\{vu,v'u' \}$. Then $H^{\dom}$ satisfies
the condition of Lemma \ref{monoton}. Therefore $H$ has the
2-Min-Max ordering.\qed

\2

The following theorem shows when MinHOM($H$) is polynomial time
solvable for the case when $H$ does contain $\vec{C}_4$ as an
induced subdigraph.

\begin{theorem} \label{C4}
Let $H$ be a semicomplete bipartite digraph. Assume that $H$ does
not contain a digraph from $\cal HFORB$ as an induced subdigraph,
but contains $\vec{C}_4$ as an induced subdigraph. Then $H$ is an
extension of a $\vec{C}_4$ and MinHOM($H$) is polynomial time
solvable.
\end{theorem}
\pf Observe that an extension $L$ of any cycle $\vec{C}_p$, $p\ge
2$, has a $p$-Min-Max ordering. Thus, MinHOM($L$) is polynomial time
solvable by Theorem \ref{mm2}. Let $C=v_1u_1v_2u_2v_1$ be an induced
4-cycle of $H$. It suffices to prove that $H$ is an extension of
$C$.

For $i=1,2$, let $M^+(v_i)= \{u \in U:\ v_iu \in A(H)  \ , \ uv_i
\not\in A(H)\}$, $M^-(v_i)=\{u \in U:\ uv_i \in A(H) \, \ v_iu
\not\in A(H)\}$, and $M(v_i)=\{u \in U:\ uv_i,v_iu \in A(H)\}$. We
have $M^+(v_1) \cap M^+(v_2) = \emptyset$ as otherwise
$H[\{v_1,v_2,u_1,u_2,u_3\}]\cong H^*$, where $u_3 \in N^+(v_1) \cap
N^+(v_2)$. We have $M^-(v_1) \cap M^-(v_2) = \emptyset$ as otherwise
$H[\{v_1,v_2,u_1,u_2,u_3\}]\cong H^{**}$, where $u_3 \in N^-(v_1)
\cap N^-(v_2)$ and $H^{**}$ is the converse of $H^*$.

We have $M(v_1) \cap (M^+(v_2) \cup M^-(v_2)) \ne \emptyset$ as
otherwise $H[\{v_1,u_1,v_2,u\}]\cong C_4'$, where $u \in  M(v_1)
\cap M^+(v_2)$ or $H[\{v_1,u_2,v_2,u\}]\cong C_4'$, where $u \in
M(v_1) \cap M^-(v_2)$. Moreover, $M(v_1) \cap M(v_2) = \emptyset$
as otherwise $H[\{v_1,u_1,v_2,u\}]\cong C_4''$, where $u \in
M(v_1) \cap M(v_2)$. The arguments above imply that
$M(v_1)=M(v_2)= \emptyset$, and $M^+(v_1)=M^-(v_2)$ and
$M^-(v_1)=M^+(v_2)$.

Similarly, we can define $M+(u_i)$, $M^-(u_i)$ and $M(u_i)$,
$i=1,2$, and prove the relations analogous to those for
$M^+(v_i)$, $M^-(v_i)$ and $M(v_i)$, $i=1,2$.

Let $v\in V-\{v_1,v_2\}$ and $u\in U-\{u_1,u_2\}$ be arbitrary.
Without loss of generality, assume that $u\in M^+(v_1)=M^-(v_2)$
and $v\in M^+(u_2)=M^-(u_1)$ (all other cases can be treated
similarly). To show that $H$ is an extension of $C$, it suffices
to prove that $v\dom u$, but $uv\not\in A(H).$ Suppose first that
$u\dom v$ and $v\dom u.$ Then $H[\{v,u,v_2,u_2\}]\cong C'_4,$ a
contradiction. Now suppose that $u\dom v$, but $vu\not\in A(H).$
Then $H[\{v,v_1,v_2,u,u_2\}]\cong H^*,$ a contradiction. Thus,
$v\dom u$, but $uv\not\in A(H)$ and we are done. \qed

\2

The three theorems of this section and the fact that $\vec{C}_4$ has
a 4-Min-Max ordering imply the following:

\begin{corollary}\label{corP}
Let $H$ be a connected semicomplete bipartite digraph not containing
a digraph from $\cal HFORB$ as an induced subdigraph. Then
MinHOM($H$) is polynomial time solvable and $H$ has a $k$-Min-Max
ordering for $k=2$ or 4.
\end{corollary}

\section{NP-hardness Cases}\label{npsec}

It is well known that the problem of finding a maximum size
independent set in an undirected graph $G$ is NP-hard. We say that
a set $I$ in a digraph $D$ is {\em independent} if no vertices in
$I$ are adjacent. Clearly, the problem of finding a maximum size
independent set in a digraph $D$ (MISD) is NP-hard.

\begin{lemma}\label{l1}
MinHOM($C_4'$) is NP-hard.
\end{lemma}
\pf  Let $H$ be isomorphic to $C_4'$ as follows: $V(H)= \{1,2,3,4
\}$ and $A(H)= \{12,21,23,34,41 \}$. Let $D$ be an arbitrary
digraph. We replace every arc $uv$ of $D$ by the digraph $G_{uv}$
with $V(G_{uv})=\{x,y, z, u,v \}$ and $A(G_{uv})= \{ux, xy,yz, vz
\}$. Let $D'$ be the obtained digraph. Define the cost function as
follows: $c_1(u)=1$, $c_2(u)=1$, $c_3(u)=0$, $c_4(u)=1$,
$c_1(v)=1$, $c_2(v)=1$, $c_3(v)=0$, $c_4(v)=1$, $c_i(t)=0$ when $t
\notin \{u,v\}$ and $i\in \{1,2,3,4\}$.

Let $h$ be a mapping from $V(G_{uv})$ to $V(H)$, and let $uv$ be
an arc in $D$. If $h(u)=h(y)=h(v)=1$ and $h(x)=h(z)=2$, then $h$
is a homomorphism. Thus, there is a homomorphism of $D'$ to $H$,
which maps all vertices of $D$ into 1 and the vertices of $D'$ not
in $D$ into 1 or 2.

Now let $f$ be a homomorphism of $G_{uv}$ to $H.$ Observe that if
$f(u)=1$, then $f(x)=2$, $f(y)$ is either 1 or 3, $f(z)$ is either
2 or 4, $f(v)$ is either 1 or 3. . Similarly if $f(u)=3$, then
$f(v) = 1$; if $f(u)=4$, then $f(v)$ is either 2 or 4; if
$f(u)=2$, then $f(v)$ is either 2 or 4.

Let $g$ be a minimum cost homomorphism of $D'$ to $H$ and let
$S=\{s\in V(D):\ g(s)=3\}.$ Notice that the cost of $g$ is
$|V(D)|-|S|$ and $S$ is an independent set by the arguments of the
previous paragraph. Thus, $S$ is an independent set of $D$ of
maximum size.

Let $I$ be a maximum size independent set in $D$. The above
arguments show that there is a homomorphism $d$ of $D'$ to $H$
such that $d(t)=3$ if $t\in I$ and $d(t)=1$ if $t\in V(D)-I$.
Notice that $d$ is a minimum cost homomorphism.

Now we can conclude that MinHOM($H$) is NP-hard since MISD is
NP-hard.\qed

\begin{lemma}\label{l3}
MinHOM($C_4''$) is NP-hard.
\end{lemma}
\pf  Let $H$ be isomorphic to $C_4''$ as follows: $V(H)= \{1,2,3,4
\}$, $A(H)= \{12,21,23,32,34,41 \}$. Let $D$ be an arbitrary
digraph. We replace every arc $uv$ of $D$ by the digraph $G_{uv}$
with $V(G_{uv})=\{x,u,v \}$ and $A(G_{uv})= \{ux, xv \}$. Let $D'$
be the obtained digraph. Define the cost function as follows:
$c_1(u)=1$, $c_2(u)=1$, $c_3(u)=1$, $c_4(u)=0$, $c_1(v)=1$,
$c_2(v)=1$, $c_3(v)=1$, $c_4(v)=0$, $c_i(x)=0$ for $i=1,2,3,4.$

Let $h$ be a mapping from $V(G_{uv})$ to $V(H)$, and let $uv$ be
an arc in $D$. If $h(u)=h(v)=1$ and $h(x)=2$, then $h$ is a
homomorphism. Thus, there is a homomorphism of $D'$ to $H$, which
maps all vertices of $D$ into 1 and the vertices of $D'$ not in
$D$ into 2.

Now let $f$ be a homomorphism of $G_{uv}$ to $H.$ Observe that if
$f(u)=1$, then $f(x)=2$ and $f(v)$ is either 1 or 3. Similarly if
$f(u)=2$, then $f(v) \in \{2,4\}$; if $f(u)=3$, then $f(v)  \in
\{1,3\}$;  if $f(u)=4$,  then  $f(v) = 2$.

Let $g$ be a minimum cost homomorphism of $D'$ to $H$ and let
$S=\{s\in V(D):\ g(s)=4\}.$ Notice that the cost of $g$ is
$|V(D)|-|S|$ and $S$ is an independent set by the arguments of the
previous paragraph. Thus, $S$ is an independent set of $D$ of
maximum size.

The rest of the proof is similar to that of Lemma \ref{l1}.\qed

\2

The following lemma was stated in \cite{gutinDO}. We give a proof
here for the sake of completeness.

\begin{lemma}\label{l4}
MinHOM($H^*$) is $NP$-hard.
\end{lemma}
\pf Let $H$ be isomorphic to $H^*$ as follows: $V(H)= \{1,2,3,4,5
\}$, $A(H)= \{12,23,34,41,15,35\}$. We replace every arc $uv$ of $D$
by the digraph $G_{uv}$ with $V(G_{uv})=
\{v_1,v_2,v_3,v_4,v_5,v_6,v_7\}$, where $u=v_6$ and $v=v_7$, and
$A(G_{uv})=\{v_1v_2,v_2v_3,v_3v_4,v_4v_1,
v_5v_6,v_5v_7,v_1v_6,v_3v_7\}$. Let $D'$ be the obtained digraph.
Define the cost function as follows: $c_1(v_6)=1$, $c_2(v_6)=1$,
$c_3(v_6)=1$, $c_4(v_6)=0$, $c_5(v_6)=1$, $c_1(v_7)=1$,
$c_2(v_{7})=1$, $c_3(v_{7})=1$, $c_4(v_{7})=0$, $c_5(v_{7})=1$, and
$c_i(v_j)=0$ for each $j \ne 6,7$.

Let $h$ be a mapping from $V(G_{uv})$ to $V(H)$, and let $uv$ be
an arc in $D$. If $h(v_i)=i$ for each $i=1,2,3,4$, $h(v_5)=1$ and
$h(u)=h(v)=5$, then $h$ is a homomorphism. Thus, there is a
homomorphism of $D'$ to $H$, which maps all vertices of $D$ into
5.

Now let $f$ be a homomorphism of $G_{uv}$ to $H.$ Observe that if
$x=f(u)=f(v)$, then $x=5$. Also, if $f(u)=5$ then $f(v) \in
\{2,4,5\}$ and if $f(v)=5$ then $f(u) \in \{2,4,5\}$.

Let $g$ be a minimum cost homomorphism of $D'$ to $H$ and let
$S=\{s\in V(D):\ g(s)=4\}.$ Notice that the cost of $g$ is
$|V(D)|-|S|$ and $S$ is an independent set by the arguments of the
previous paragraph. Thus, $S$ is an independent set of $D$ of
maximum size.

The rest of the proof is similar to that of Lemma \ref{l1}. \qed

\2

\begin{lemma} \label{l6}
MinHOM$(N_1)$ is NP-hard.
\end{lemma}
\pf We shall reduce the maximum independent set problem to
$MinHOM(N_1)$. Let $H$ be the following digraph isomorphic to
$N_1$: $V(H)= \{1,2,3,4,5,6\}$,
$$A(H)=\{12,21, 34,43, 56,65, 23, 25,14,16,54,36\}.$$
Let $D$ be an arbitrary digraph. We replace every arc $uv$ of $D$
with the digraph $G_{uv}$ with $V(G_{uv})=\{x,y, z, u,v \}$ and
$A(G_{uv})= \{ux, vz, xy,yz \}$.  Consider the following cost
function: $c_2(u)=c_2(v)=1$, $c_6(u)=c_6(v)=0$,
$c_i(u)=c_i(v)=2M+1$ for $i \ne 2,6$ and $c_6(y)=2M+1$, where
$M=|V(D)|$. In all remaining cases the cost is zero. Let $D'$ be
the obtained digraph, let $f$ be a mapping from $V(D')$ to $V(H)$,
and let $uv$ be an arc in $D$.

Assume that $f(u)=f(v)=2$. Then with $f(x)=f(z)=1$ and $f(y)=2$,
we obtain a homomorphism from $G_{uv}$ to $H$ of cost $2$. This
implies there is a homomorphism of $D'$ to $H$ of cost $2M<2M+1$,
and, thus, every vertex of $D$ in $D'$ must be colored either 2 or
6 in any minimum cost homomorphism of $D'$ to $H$. Let $f$ be a
homomorphism of $D'$ to $H$ and let us consider the remaining
options for coloring the vertices of $D$ in $D'$.

Assume that $f(v)=6$ and $f(u)=2$. Then with $f(z)=5$, $f(y)=2$
and $f(x)=1$, we obtain a homomorphism from $G_{uv}$ to $H$ of
cost $1$. Assume that $f(v)=2$ and $f(u)=6$. Then with $f(x)=5$,
$f(y)=4$ and $f(z)=3$, we obtain a homomorphism from $G_{uv}$ to
$H$ of cost $1$. Note that if $f(u)=f(v)=6$, then $f(x)=f(z)=5$
and $f(y)=6$. Then the cost of $f$ will be at least $2M+1$
implying we cannot color both vertices $u$ and $v$ in color 6 in
any minimum cost homomorphism of $D'$ to $H$.

Now let $f$ be a minimum cost homomorphism,  let $S$ be the
vertices of $D$ in $D'$ colored 6 and $T=V(D)-S$. Recall that the
vertices of $T$ are colored 2. Notice that $S$ is an independent
set and the cost of $f$ equals $|T|$.

The rest of the proof is similar to that of Lemma \ref{l1}.\qed

\begin{lemma} \label{l7}
MinHOM($N_2$) is NP-hard.
\end{lemma}
We shall reduce the maximum independent set problem to
MinHOM($N_2$). Let $H$ be the following digraph isomorphic to
$N_2$ : $V(H)= \{1,2,3,4,5,6\}$,
$$A(H)=\{12, 34,43, 56,65, 23, 25,14,16,54,36\}.$$
Let $D$ be an arbitrary digraph. We replace every arc $uv$ of $D$
by the digraph $G_{uv}$ with $V(G_{uv})=\{x,y, z, u,v \}$ and
$A(G_{uv})= \{ux,vz, xy, zy \}$. We introduce the following cost
function: $c_1(u)=c_1(v)=1$, $c_5(u)=c_5(v)=0$,
$c_i(u)=c_i(v)=2M+1$ for $i \ne 1,5$ and $c_4(x)=2M+1$ and
$c_6(z)=2M+1$, where $M=|V(D)|$. In any other cases the cost is
zero. Let $D'$ be the obtained digraph, let $f$ be a mapping from
$V(D')$ to $V(H)$, and let $uv$ be an arc in $D$.

Assume that $f(u)=f(v)=1$. With $f(x)=f(z)=2$ and $f(y)=3$ with
obtain a homomorphism from $H'$ to $H$ with cost $2$. Thus, there
is a homomorphism of $D'$ to $H$ of cost at most $2M$ (assign all
vertices of $D$ in $D'$ color 2) and no vertex of $D$ in $D'$ must
not be assigned any color other than 1 and 5.  Let $f$ be a
homomorphism of $D'$ to $H$ and let us consider the remaining
options for coloring the vertices of $D$ in $D'$.

Assume that $f(v)=1$ and $f(u)=5$. With $f(z)=2$, $f(x)=6$ and
$f(y)=5$, we obtain a homomorphism of $G_{uv}$ to $H$ of cost $1$.
Assume that $f(v)=5$ and $f(u)=1$. With $f(x)=2$, $f(z)=4$ and
$f(y)=3$,  we obtain a homomorphism of $G_{uv}$ to $H$ of cost
$1$. Note that if $f(u)=f(v)=5$, then $f(x) \in \{4,6\}$ and $f(z)
\in \{4,6\}$. Thus, $f$ has cost at least $2M+1$ implying that a
minimum cost homomorphism of $D'$ to $H$ does not assign adjacent
vertices of $D$ color 5 (in $D'$).

Now let $f$ be a minimum cost homomorphism,  let $S$ be the
vertices of $D$ in $D'$ colored 5 and $T=V(D)-S$. Recall that the
vertices of $T$ are colored 1. Notice that $S$ is an independent
set and the cost of $f$ equals $|T|$.

The rest of the proof is similar to that of Lemma \ref{l1}.\qed

\begin{corollary} \label{corNP}
MinHOM($H$) is NP-hard for every $H \in \cal HFORB$.
\end{corollary}
\pf If $H$ is isomorphic to $C_4'$, $C_4''$, $H^*$, $N_1$ or $N_2$
or the converse of one of the five digraphs, then MinHOM($H$) is
NP-hard due to the lemmas of this section and the simple fact that
if MinHOM($H$) is NP-hard and $H'$ is the converse of $H$ then
MinHOM($H'$) is NP-hard as well.

Let $\cal B$ be the set consisting of the following bipartite
graphs: bipartite claw, bipartite net, bipartite tent and every even
cycle with at least 6 vertices. If $UN(H^s)$, where $s\in \{\dom,
\ldom\}$, is isomorphic to a graph in $\cal B$, then MinHOM($H$) is
NP-hard due to Theorem \ref{unddich} and the transformation from a
bipartite undirected graph to a semicomplete bipartite digraph
described in the last paragraph of subsection `Minimum Cost
Homomorphisms' of Section \ref{introsec}. If $UN(H^{\edom})$ is
isomorphic to a graph in $\cal B$, then MinHOM($H$) is NP-hard as,
for each bipartite undirected graph $L$, MinHOM($L$) is equivalent
to MinHOM($L^+$), where $L^+$ is the digraph obtained from $L$ by
replacing every edge $xy$ with two arcs $xy$ and $yx.$\qed

\section{Dichotomy for semicomplete multipartite
digraphs}\label{gdsec}

A digraph $D$ is called {\em semicomplete $k$-partite} if $D$ can be
obtained from a complete $k$-partite (undirected) graph $G$ by
replacing every edge $xy$ of $G$ by either the arc $xy$ or the arc
$yx$ or the pair $xy,yx$ of arcs. Let $TT_p$ denote the acyclic
tournament on $p\ge 1$ vertices. Let $p\ge 3$ and let $TT_p^-$ be a
digraph obtained from $TT_p$ by deleting the arc from the vertex of
in-degree zero to the vertex of out-degree zero. Combining the main
result of this paper with the main result of \cite{gutinDO}, we
obtain the following:

\begin{theorem}
Let $H$ be a semicomplete $k$-partite digraph. If $k=2$ and $H$ does
not contain a digraph from $\cal HFORB$ as an induced subdigraph or
if $k\ge 3$ and $H$ is an extension of either $TT_k$ or $TT^-_{k+1}$
or $\vec{C}_3$, then MinHOM($H$) is polynomial time solvable.
Otherwise, MinHOM($H$) is NP-hard.
\end{theorem}

\section{Further Research}\label{frsec}

In the case of undirected graphs $H$, the well-known theorem of Hell
and Ne\v{s}et\v{r}il \cite{hellJCT48} on the homomorphism problem
implies that MinHOM($H$) is NP-hard for each non-bipartite graph
$H$. The authors of \cite{gutinEJC} obtained a complete dichotomy of
the computational complexity of MinHOM($H$) when $H$ is undirected.
The dichotomy obtained in this paper significantly extends the
dichotomy of \cite{gutinEJC}. This indicates that the problem of
obtaining a dichotomy for the computational complexity of
MinHOM($H$) when $H$ is a bipartite digraph is a very difficult
problem. Note that MinHOM($H$) is polynomial-time solvable for some
non-bipartite digraphs, for example, for acyclic tournaments
\cite{gutinDO}. Thus, a dichotomy for bipartite directed case does
not coincide with a dichotomy for the general directed case. The
problem of obtaining dichotomy for both cases is a very interesting
open problem.

\2

{\bf Acknowledgements} We are grateful to Pavol Hell for several
useful discussions. Research of the first author was supported in
part by the IST Programme of the European Community, under the
PASCAL Network of Excellence, IST-2002-506778.

{\small

\end{document}